# Detection of Ship Wakes in SAR Imagery Using Cauchy Regularisation

Tianqi Yang, Oktay Karakuş, Alin Achim

**Abstract**

Ship wake detection is of great importance in the characterisation of synthetic aperture radar (SAR) images of the ocean surface since wakes usually carry essential information about vessels. Most detection methods exploit the linear characteristics of the ship wakes and transform the lines in the spatial domain into bright or dark points in a transform domain, such as the Radon or Hough transforms. This paper proposes an innovative ship wake detection method based on sparse regularisation to obtain the Radon transform of the SAR image, in which the linear features are enhanced. The corresponding cost function utilizes the Cauchy prior, and on this basis, the Cauchy proximal operator is proposed. A Bayesian method, the Moreau-Yoshida unadjusted Langevin algorithm (MYULA), which is computationally efficient and robust is used to estimate the image in the transform domain by minimizing the negative log-posterior distribution. The detection accuracy of the Cauchy prior based approach is 86.7%, which is demonstrated by experiments over six COSMO-SkyMed images.

## 1. INTRODUCTION

Ship wakes are significant features frequently found in synthetic aperture radar (SAR) images of the ocean surface. Since they provide key information on the vessels that can be used in velocity estimation, tracking, classification, and so on, detecting ship wakes turns out to be a timely and important research topic.

Considering the possibility to model ship wakes as linear features, ship wake detection approaches generally exploit line detection methods, such as the Hough and Radon transforms. Thanks to its low computational cost compared to the Hough transform, the Radon transform has been initially utilized by Murphy [1] in ship wake detection. Rey et al. [2] have combined Wiener filtering with the Radon transform to enhance information in the Radon domain and improve peak detectability. Tunaley has used a method to limit the search range in the Radon domain in [3] and improved the detection performance. In an application focused on ship velocity estimation, Zilman et al. [4] have performed wake detection by enhancing the information in the Radon domain. Courmontagne [5] has combined Radon transform with stochastic matched filtering, whilst Kuo and Chen [6] have utilized wavelet correlators for wake detection.

In recent work on wake detection, which processes observed SAR images directly without performing any enhancement step, Graziano et al. [7-10] have proposed the idea of using ship-centred-masked images and limiting the searching area in the Radon domain within two sine waves. Karakuş et al. [11, 12] have proposed a novel technique for ship wake detection in SAR images based on sparse regularization whereby the generalized minimax concave (GMC) penalty of Selesnick [13] has first been introduced in a 2-D image

This work was supported by the Engineering and Physical Sciences Research Council (EPSRC) under grant EP/R009260/1 (AssenSAR).
Tianqi Yang, Oktay Karakuş, Alin Achim are with the Visual Information Lab, University of Bristol, Bristol BS1 5DD, U.K.
(e-mail: yang_tq@outlook.com; o.karakus@bristol.ac.uk; Alin.Achim@bristol.ac.uk)

enhancement study. In [11], the performance of the GMC based method has been demonstrated with at least 10% accuracy gain over various datasets compared to common regularisers, such as $\mathrm{L}_1$, $\mathrm{L}_\mathrm{p}$ and TV norms.

The Cauchy distribution is a member of the α-stable distribution family and known for its ability to model heavy-tailed data. As a prior, it is sparsity-enforcing similar to its generalized-Gaussian counterpart, the Laplace distribution [14] (i.e. the $\mathrm{L}_1$ norm) and it has generally been utilized in despeckling studies by modelling subband coefficients of contourlet [15], and wavelet [16, 17] transforms. Apart from its applications in denoising studies, it has otherwise very rarely been used in imaging inverse problems, due to the absence of a proximal operator for the Cauchy prior, which would make it applicable in basic proximal splitting algorithms such as forward backward (FB).

In this paper, we exploit the flexibility of the method proposed in [11] and improve it by utilizing Cauchy distribution as prior instead of the GMC. As the proposed methodology incorporates the Cauchy distribution as the regularization term, we further derive the proximal operator for it, which, we believe, will make Cauchy regularisation more applicable in various imaging inverse problems. Moreover, the solution to the Cauchy regularized inverse problem for wake detection is found through a Bayesian methodology, namely Moreau-Yoshida unadjusted Langevin algorithm (MYULA), which has high computational efficiency and robustness. In the experimental analysis, SAR images of the sea surface acquired by the COSMO-SkyMed satellite are used for ship wake detection and performance analysis of the proposed method is performed in comparison to the GMC regularisation in [11] and the detection method in Graziano et. al. [7] .

The rest of the paper is structured as follows: the background is presented in Section 2, including the image formation model for ship wakes identification and maximum a-posteriori (MAP) image reconstruction via MYULA. In Section 3, MYULA implementation in solving the Cauchy prior based inverse problem and the derivation of the Cauchy proximal operator are discussed. The experimental results are shown in Section 4, and Section 5 presents a brief conclusion of the paper.

## 2. THEORETICAL PRELIMINARIES

### *2.1. Image Formation Model*

Wake structures can generally be divided into four categories: a turbulent wake, Kelvin wake, narrow-V wake, and internal wave wakes, which are shown in Fig. 1 [18]. Since internal waves are observed in shallow waters and are generally less common in SAR imagery, in this paper, we carry out ship wake detection by focusing on the first three categories: (1) turbulent wake which is the central dark streak, (2) two bright arms of Narrow-V wake bounding around the central line, and (3) two Kelvin arms that are located at both sides of the turbulent wake within a half angle of 19.5°.

Ship wakes generally appear as bright and/or dark streaks over the sea surface in SAR images, and thus wake detection algorithms involve detecting linear features in noisy background. Mathematically, once we model ship wakes as lines, the SAR image formation model can be expressed in terms of its inverse Radon transform as

$$Y = CX + N \tag{1}$$

where $Y$ is the $M \times M$ SAR image, $X(r, \theta)$ is the image in Radon domain, $N$ refers to the additive noise and $C = R^{-1}$ represents the inverse Radon transform. $X(r, \theta)$ represents lines as a distance $r$ from the center of $Y$ and an orientation $\theta$ from the horizontal axis of $Y$. Discrete operators $R$ and $C$ are used as in [19].

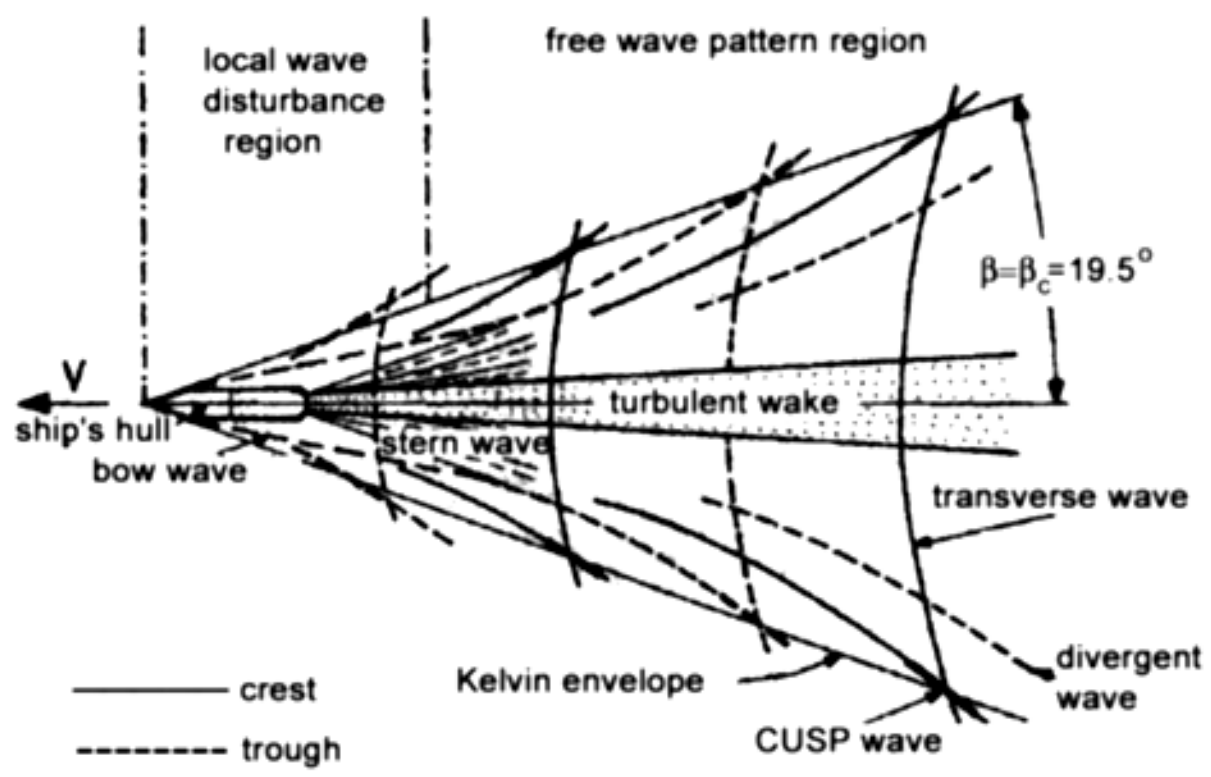

**Fig. 1.** Typical ship wake patterns [18]

### *2.2. MAP Image Reconstruction*

We consider the ship wake detection inverse problem as reconstructing the Radon image $X$ from the observed SAR image $Y$. Following Bayes' rule, incorporating a prior distribution in conjunction with the observed statistical model produces the knowledge on $X$ given $Y$, namely the posterior distribution $p(X|Y)$ as:

$$p(X|Y) = \frac{p(Y|X)p(X)}{\int p(Y|X)p(X)dX} \tag{2}$$

where the denominator $\int p(Y|X)p(X)dX$ is basically the marginal likelihood $p(Y)$ which is independent from $X$ and assumed to be constant. Hence, the unnormalised posterior is defined by

$$p(X|Y) \propto p(Y|X)p(X). \tag{3}$$

In the convex case, the posterior distribution is assumed to be log-concave, $p(X|Y) \propto exp\{-F(X)\}$, where $F(X)$ is denoted as a convex function. In modern statistical image processing many approaches rely heavily on the maximum a-posterior (MAP) estimator:

$$\hat{X}_{MAP} = arg\max_{X} p(X|Y) = arg\min_{X} F(X) \tag{4}$$

which can be computed efficiently in most cases by using proximal optimization algorithms, even in very large problems [20].

Under the assumption of an independent and identically distributed (iid) standard normal noise case, the likelihood distribution $p(Y|X)$ can be expressed as

$$p(Y|X) \propto exp\{-\|Y - CX\|_2^2\}. \tag{5}$$

We consider models in which the cost function is expressed as a summation of two functions as

$$F(X) \propto f(x) + g(x) \tag{6}$$

where $f(x) = \|Y - CX\|_2^2$ refers to the standard normal likelihood, and $g(x) = -log\,(p(X))$, in which $p(X)$ refers to the prior knowledge on $X$.

### *2.3. MYULA*

In this paper, we employ a computationally efficient, theoretically sound, and robust Markov chain Monte Carlo (MCMC) algorithm, i.e. MYULA, to solve the optimization problem given in (4). MYULA has been successfully applied to solve imaging inverse problems including image deconvolution and tomographic reconstruction in [20], and radio interferometric image reconstruction in [21]. Furthermore in [20, 21], MYULA has been used to quantify uncertainty via posterior credible sets and hypothesis testing, which demonstrate that MYULA is able to provide reliable approximation with lower computational cost than the companion sampling mechanism proximal Metropolis-adjusted Langevin algorithm (Px-MALA) [21].

Denote the posterior $p(X|Y)$ as $\pi$. For MYULA, the samples from $\pi$ can be obtained using the Langevin diffusion [21], but with the approximation shown in (7) where the Moreau-Yoshida envelope $g^{\omega}$ of the non-smooth term $g$ is included. Then the posterior can be approximated as

$$\pi_{\omega}(x) = \frac{exp\{-g^{\omega}(x)-f(x)\}}{\int exp\{-g^{\omega}(x)-f(x)\}dx} \tag{7}$$

where $\nabla log \pi_{\omega}$ is Lipchitz continuous, hence the Langevin diffusion associated with $\pi_{\omega}$ is well posed and a Markov chain with good convergence properties is obtained.

Given $\omega > 0$ and the step size $\delta > 0$, the corresponding MYULA sampling mechanism can be written as [20]

$$X^{(i+1)} = \left(1 - \frac{\delta}{\omega}\right) X^{(i)} - \delta \nabla f\left(X^{(i)}\right) + \frac{\delta}{\omega} prox_g^{\omega}\left(X^{(i)}\right) + \sqrt{2\delta} Z^{(i+1)} \tag{8}$$

where $Z^{(i)}$ is a sequence of iid standard Gaussian random variables with the same dimension as $X^{(i)}$ [20] and $prox_g^{\omega}(\cdot)$ is the Moreau proximal operator of the function $g(\cdot)$.

## 3. THE PROPOSED METHOD

### *3.1. Cauchy prior and Cauchy proximal operator*

In this study, we propose the use of the Cauchy distribution to solve imaging inverse problems by defining its proximal operator in a closed-form solution. Cauchy distribution is one of the special members of the α-stable distribution family which is known to be heavy-tailed and promote sparseness in various application areas.

Contrary to the general α-stable family, it has a closed-form probability density function, which is defined by [22]

$$p(X) \propto \frac{\gamma}{\gamma^2 + X^2} \tag{9}$$

where $\gamma$ is the dispersion (scale) parameter, which controls the spread of the distribution.

Consequently, by replacing $p(X)$ with the Cauchy prior in (9), the minimization with Cauchy regularization becomes

$$\hat{X}_{Cauchy} = arg \min_{x} \|Y - CX\|_2^2 - log\left(\frac{\gamma}{\gamma^2 + X^2}\right). \tag{10}$$

As mentioned above in (8), MYULA necessitates a proximal operator for the associated prior. The proximal operator of any function $g(\cdot)$ with $\omega > 0$ is defined as

$$prox_g^{\omega}(x) = arg \min_{u} \left[g(u) + \frac{\|u-x\|^2}{2\omega}\right]. \tag{11}$$

For a Cauchy prior, we recall $g(x) = -log\ (p(X))$, then the Cauchy proximal operator is obtained as

$$prox_g^{\omega}(x) = arg \min_{u} \left[- log\left(\frac{\gamma}{\gamma^2 + u^2}\right) + \frac{\|u-x\|^2}{2\omega}\right] \tag{12}$$

The solution to this minimization problem can be obtained by taking the first derivative of the cost function in (12) in terms of $u$ and setting it to zero. Then, we get

$$u^3 - xu^2 + (\gamma^2 + 2\omega)u - x\gamma^2 = 0. \quad (13)$$

Wan et al. in [22] proposed a MAP solution to the denoising problem of a Cauchy signal under Gaussian noise, and defined this solution as "Cauchy shrinkage". Similarly, the minimization problem in (12) can be solved with the same approach as in [22], however with different parameterization.

Hence, following [22], the solution to the cubic function given in (13) can be obtained through Cardano's method and the Cauchy proximal operator can be obtained as

$$prox_g^{\omega}(x) = \frac{x}{3} + s + t \quad (14)$$

where $s$ and $t$ are determined by $x$ and $\omega$, which are iteratively updated at each iteration together with a constant value $\gamma$, and are defined as

$$s = \sqrt[3]{\frac{q}{2} + dd} \quad (15)$$

$$t = \sqrt[3]{\frac{p}{2} - dd} \quad (16)$$

$$dd = \sqrt{\frac{p^3}{27} + \frac{q^2}{4}} \quad (17)$$

$$p = \gamma^2 + 2\omega - \frac{x^2}{3} \quad (18)$$

$$q = x\gamma^2 + \frac{2x^3}{27} - \frac{x}{3}(\gamma^2 + 2\omega) \quad (19)$$

Although the cost function in (10) is non-convex, the use of a Cauchy prior guarantees that the approximate full posterior $\pi^{\omega}$ defined in (7) remains proper. Consequently, MYULA can still be used in our case [20].

*3.2. MYULA for Cauchy Regularized Cost Function*

Since we have the proximal operator definition for a Cauchy prior, the minimization problem given in (10) can be solved via MYULA which as shown in Algorithm I. The algorithm stops when one of the following conditions is satisfied.

i) When the maximum number of iterations is reached. In this study we set $MaxIter = 200$.

ii) The error term reaches a small value of $10^{-3}$. For each iteration $i$, this term is calculated as

$$\epsilon^{(i)} = \frac{\|X^{(i)} - X^{(i-1)}\|}{\|X^{(i-1)}\|} \quad (20)$$

We observed that in order to obtain a point estimate, small number of iterations, e.g. 150 or 200, are enough. However, please note that in a fully Bayesian analysis such as uncertainty quantification, Algorithm I should be run longer in order to obtain a proper posterior distribution for each pixel.

The choice of MYULA parameters, $\delta$ and $\omega$ given in Algorithm I are related to the Lipschitz constant $L$. According to Theorem 2 in [20], $\delta$ should take values in the interval $(0, \omega/(L \cdot \omega + 1)]$ to ensure the stability of the Euler-Maruyama discretization. The value of $\delta$ is closely related to the bias-variance trade-off. Specifically, small $\delta$ values can produce low asymptotic bias, but create a Markov chain that converges slowly and requires large number of iterations to produce stable estimates. Conversely, a large value of $\omega$ speeds up convergence at the expense of asymptotic bias. Thence, we use $\omega = 1/4L$ and $\delta \in [1/25L, 1/10L]$ in our experiments.

---

**Algorithm I** MYULA for Cauchy regularized cost function

**Input:** SAR image $Y$, $\gamma \in [0.0001, 0.1]$
**Output:** Radon image $X$
**Set:** $\delta = 1/25L$, $\omega = 1/4L$
**do**
$Z^{(i+1)} \sim N(0, \mathbb{I}_d)$
$X^{(i+1)} = \left(1 - \frac{\delta}{\omega}\right) X^{(i)} - \delta \nabla f\left(X^{(i)}\right) + \frac{\delta}{\omega} prox_g^{\omega}\left(X^{(i)}\right) + \sqrt{2\delta} Z^{(i+1)}$
**while** $\epsilon^{(i)} > 10^{-3}$ or $i < MaxIter$

---

*3.3. Wake Detection*

Upon obtaining the solution to the inverse problem, the reconstructed Radon image is used to detect ship wakes for the corresponding ship-centred-masked SAR image. The method used in this paper is the one proposed in [11]. In particular, we first limit the searching area between two sine waves as in [7, 11]. Then the turbulent and one narrow-V wake pair satisfying the angular condition, i.e. that they are no more than 4° apart, is detected. Kelvin wake's detection is similar to the above process, except that the angular distance is 19.5°.

Thereby, confirmation of the candidate half-lines is achieved by using as measure an index, $F_I$. This is interpreted as a measure of the ratio between the mean value over the un-confirmed wake, $\bar{I}_w$, and the mean intensity of the image, $\bar{I}$ [11]:

$$F_I = \bar{I}_w / \bar{I} - 1 \tag{21}$$

Assuming a margin of 10%, after a trial-error procedure an index $F_I > 0.1$ is obtained for both the narrow-V wake and Kelvin wake. Therefore, half-lines satisfying $F_I \geq 0$ for turbulent wakes, and $F_I \leq 0.1$ for narrow-V and Kelvin wakes are confirmed, whereas those not following these rules are discarded. For further details on the part of the method please see [7, 11].

**Table 1.** Visible wakes in used image dataset
(1 means visible and 0 represents invisible)

| Image | Turbulent | 1st Narrow | 2nd Narrow | 1st Kelvin | 2nd Kelvin |
|---|---|---|---|---|---|
| CSM_1 | 1 | 1 | 0 | 0 | 0 |
| CSM_2 | 1 | 1 | 0 | 0 | 0 |
| CSM_3 | 1 | 1 | 0 | 1 | 0 |
| CSM_4 | 1 | 1 | 0 | 1 | 0 |
| CSM_5 | 1 | 1 | 0 | 0 | 0 |
| CSM_6 | 1 | 1 | 0 | 0 | 0 |

**Table 2.** Detection results over 6 COSMO-SkyMed images
($\omega = 1/4L, \delta = 1/25L$)

| | TP | TN | FP | FN | %Accuracy |
|---|---|---|---|---|---|
| Cauchy | 40.0% | 46.7% | 6.7% | 6.7% | 86.7% |
| GMC [11] | 36.7% | 40% | 20% | 3.3% | 76.7% |
| Graziano [7] | 33.3% | 36.7% | 16.7% | 13.3% | 70.0% |

## 4. EXPERIMENTAL RESULTS

In order to test the detection performance of the proposed method, we used 6 COSMO-SkyMed images. Table 1 presents the ship wake information identified by visual inspection for all 6 images, where 1 means visible and 0 represents invisible wakes. In Table 1, there are 14 detectable wakes (6 turbulent, 6 Narrow-V and 2 Kelvin) out of 30 wakes. The performance evaluation is based on the percentage of correctly detected/discarded ship wakes.

In the experimental analysis, we compared the wake detection performance of the proposed method to the two state-of-the-art methods, which are proposed in [11] for GMC regularisation, and by Graziano et. al. in [7], respectively for all 6 SAR images utilised in this paper. As described in Section 3, in the proposed method, the value of $\omega$ is fixed to $1/4L$, and a value of $\delta$ in the range of $[1/25L, 1/10L]$ is tested, from which the value with the best performance is selected. Evaluating the detection accuracy consists of two components, i.e. the percentage of correctly confirmed (true positive (TP)) visible wakes as well as the percentage of correctly discarded (true negative (TN)) for invisible wakes.

Table 2 presents the detection performance over 6 COSMO images for all methods. By examining TP and TN values in Table 2, we can see that the detection results in the Cauchy case are 40.0% and 46.7% for TP and TN, respectively, whilst the GMC has 36.7% and 40.0%. Compared to the regularisation based methods the method of Graziano et. al. [7] performs poorer since it does not perform any image enhancement methodology. This fact obviously states that the regularisation-based wake detection methodology proposed in [11] is crucial for wake detection by having 16.7% accuracy gain over all images in this paper. The choice of Cauchy prior is also another important factor, which is shown in this paper. The overall accuracy for all 6 images is 10% higher than the GMC based prior proposed in [11].

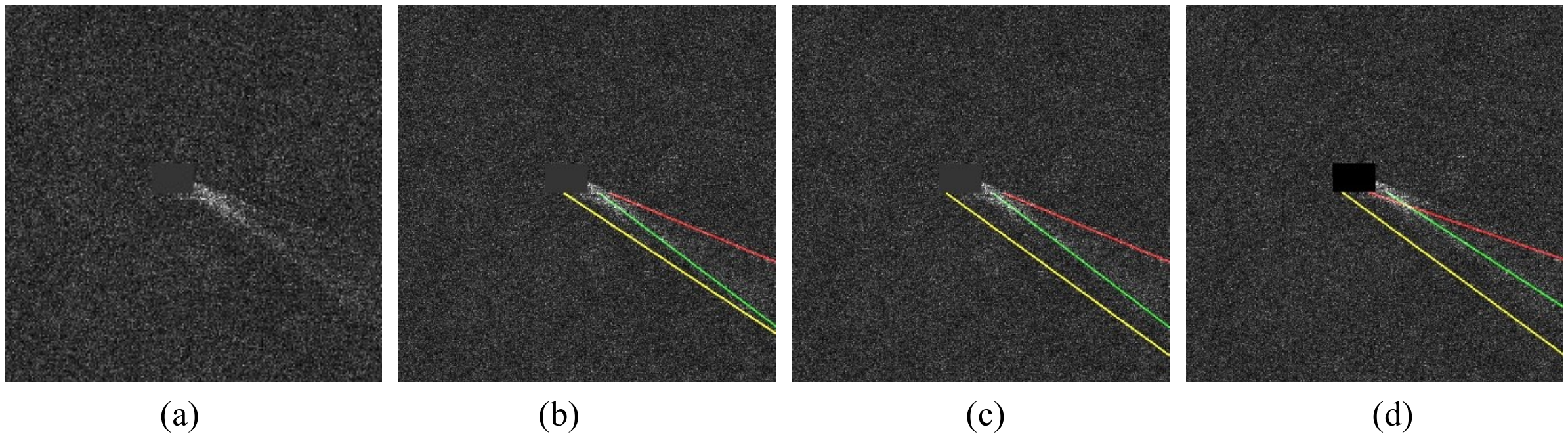

(a) (b) (c) (d)

**Fig. 2.** Ship wake detection results for CSM_4. (a) Masked ship-centred image. Detected lines (b), (c) and (d) for the proposed method, [11] and [7], respectively. Yellow, green and red lines refer to the turbulent, narrow-V and Kelvin wakes, respectively.

Furthermore, visual evaluation for all methods can be performed by observing Fig. 2. From the spatial domain, it can be seen that all the methods detect three visible wakes. However, the Cauchy based approach is able to locate the turbulent wake closer to the true location compared with the GMC detection as well as the method of Graziano et. al, which demonstrates the higher accuracy of the proposed method.

## 5. CONCLUSIONS

In this paper, we proposed a new method for detecting ship wakes in SAR images of the sea surface based on solving an inverse problem. The Cauchy prior was used in the definition of the cost function, and the corresponding solution was found by using an efficient and robust Bayesian method, i.e. MYULA. A Cauchy proximal operator was introduced for the first time, in conjunction with proximal splitting (or proximal MCMC) algorithms and applied to the ship wake detection in SAR images. In the experiments, 6 COSMO-SkyMed satellite images were used to test the performance of the proposed method. The superiority of the proposed method was demonstrated through the comparison with the GMC prior based method [11] as well as the method of Graziano et. al. [7] with at least 10% accuracy gain over the dataset.

## 6. ACKNOWLEDGEMENT

We are grateful to the UK Satellite Applications Catapult for providing the COSMO-SkyMed datasets employed in this study.

## 7. REFERENCES

[1] L. M. Murphy, "Linear feature detection and enhancement in noisy images via the Radon transform," Pattern Recognition Letters, vol. 4, no. 4, pp. 279-284, 1986/09/01.

[2] M. T. Rey, J. K. Tunaley, J. T. Folinsbee, P. A. Jahans, J. A. Dixon, and M. R. Vant, "Application of Radon Transform Techniques to Wake Detection In Seasat-A SAR Images," IEEE Transactions on Geoscience and Remote Sensing, vol. 28, no. 4, pp. 553-560, 1990.

[3] J. K. E. Tunaley, "The estimation of ship velocity from SAR imagery," in IGARSS 2003. 2003 IEEE International Geoscience and Remote Sensing Symposium. Proceedings (IEEE Cat. No.03CH37477), 21-25 July 2003, vol. 1, pp. 191-193 vol.1.

[4] G. Zilman, A. Zapolski, and M. Marom, "The speed, and beam of a ship 'From its wake's SAR images," Ieee Transactions on Geoscience and Remote Sensing, vol. 42, no. 10, pp. 2335-2343, Oct 2004.

[5] P. Courmontagne, "An improvement of ship wake detection based on the Radon transform," Signal Processing, vol. 85, no. 8, pp. 1634-1654, 2005.

[6] J. M. Kuo and K.-S. Chen, "The application of wavelets correlator for ship wake detection in SAR images," IEEE Transactions on Geoscience and Remote Sensing, vol. 41, no. 6, pp. 1506-1511, 2003.

[7] M. D. Graziano, M. D'Errico, and G. Rufino, "Wake Component Detection in X-Band SAR Images for Ship Heading and Velocity Estimation," Remote Sensing, vol. 8, no. 6, Jun 2016, Art no. 498.

[8] M. D. Graziano, M. D'Errico, and G. Rufino, "Ship heading and velocity analysis by wake detection in SAR images," Acta Astronautica, vol. 128, pp. 72-82, 2016/11/01.

[9] M. D. Graziano, G. Rufino, and M. D'Errico, Wake-based ship route estimation in high-resolution SAR images (SPIE Remote Sensing). SPIE, 2014.

[10] M. Graziano, M. Grasso, and M. D'Errico, Performance Analysis of Ship Wake Detection on Sentinel-1 SAR Images. 2017, p. 1107.

[11] Karakuş, Oktay & Rizaev, Igor & Achim, Alin. (2019). Ship Wake Detection in SAR Images via Sparse Regularization. IEEE Transactions on Geoscience and Remote Sensing. PP. 1-13. 10.1109/TGRS.2019.2947360.

[12] O. Karakuş and A. Achim, "Ship Wake Detection in X-band SAR Images Using Sparse GMC Regularization," in ICASSP2019 - 2019 IEEE International Conference on Acoustics, Speech and Signal Processing (ICASSP), 12-17 May 2019, pp. 2182-2186.

[13] I. Selesnick, "Sparse regularization via convex analysis." IEEE Transactions on Signal Processing 65.17 (2017):

4481-4494.
[14] A. Mohammad-Djafari, "Bayesian approach with prior models which enforce sparsity in signal and image processing," EURASIP Journal on Advances in Signal Processing, journal article vol. 2012, no. 1, p. 52, March 01 2012.
[15] H. Sadreazami, M. O. Ahmad, and M. N. S. Swamy, "Ultrasound image despeckling in the contourlet domain using the Cauchy prior," in 2016 IEEE International Symposium on Circuits and Systems (ISCAS), 22-25 May 2016, pp. 33-36.
[16] C. Guozhong and L. Xingzhao, "Wavelet-based SAR image despeckling using Cauchy pdf modeling," in 2008 IEEE Radar Conference, 26-30 May 2008, pp. 1-5.
[17] A. Achim and E. E. Kuruoglu, "Image denoising using bivariate α-stable distributions in the complex wavelet domain," IEEE Signal Processing Letters, vol. 12, no. 1, pp. 17-20, 2004.
[18] W. G. Pichel, P. C. Colon, C. C. Wackerman, and K. S. Friedman. Ship and Wake Detection [Online] Available: http://www.sarusersmanual.com/
[19] B. T. Kelley and V. K. Madisetti, "The fast discrete Radon transform. I. Theory," IEEE Transactions on Image Processing, vol. 2, no. 3, pp. 382-400, 1993.
[20] A. Durmus, E. Moulines, and M. Pereyra, "Efficient Bayesian computation by proximal Markov chain Monte Carlo: when Langevin meets Moreau," SIAM Journal on Imaging Sciences, vol. 11, no. 1, pp. 473-506, 2018.
[21] X. H. Cai, M. Pereyra, and J. D. McEwen, "Uncertainty quantification for radio interferometric imaging - I. Proximal MCMC methods," Monthly Notices of the Royal Astronomical Society, vol. 480, no. 3, pp. 4154-4169, Nov 2018.
[22] T. Wan, N. Canagarajah and A. Achim, "Segmentation of noisy colour images using Cauchy distribution in the complex wavelet domain," in *IET Image Processing*, vol. 5, no. 2, pp. 159-170, March 2011.